 \documentclass[aps,prb,reprint,showpacs,superscriptaddress]{revtex4-1}

\usepackage{mathrsfs}
\usepackage{graphicx,epsfig}
\usepackage{epstopdf}
\usepackage{rotating}
\usepackage{subfigure}
\usepackage{color}
\usepackage{amsfonts}
\usepackage{tikz}
\pdfoutput=1

\begin{document}

\title{Vortical versus skyrmionic states in mesoscopic \emph{p}-wave superconductors}

\author{V.\ Fern\'andez Becerra}
\affiliation{Departement Fysica, Universiteit Antwerpen,
Groenenborgerlaan 171, B-2020 Antwerpen, Belgium}
\author{E.\ Sardella}
\affiliation{UNESP-Universidade Estadual Paulista, Departamento de
F\'{\i}sica, Faculdade de Ci\^encias, Caixa Postal 473, 17033-360,
Bauru-SP, Brazil}
\affiliation{UNESP-Universidade Estadual Paulista,
IPMet-Instituto de Pesquisas Metereol\'ogicas, CEP 17048-699 Bauru,
SP, Brazil}
\author{F.\ M.\ Peeters}
\affiliation{Departement Fysica, Universiteit Antwerpen,
Groenenborgerlaan 171, B-2020 Antwerpen, Belgium}
\author{M.\ V.\ Milo\v{s}evi\'c}
\email{Milorad.Milosevic@uantwerpen.be}
\affiliation{Departement Fysica, Universiteit Antwerpen,
Groenenborgerlaan 171, B-2020 Antwerpen, Belgium}

\date{\today}

\begin{abstract}
We investigate the superconducting states that arise as a
consequence of mesoscopic confinement and a multi-component order
parameter in the Ginzburg-Landau model for $p\,$-wave
superconductivity. Conventional vortices,
but also half-quantum vortices and skyrmions are found as
the applied magnetic field and the anisotropy parameters of the 
Fermi surface are varied. The solutions
are well differentiated by a topological charge that for skyrmions
is given by the Hopf invariant and for vortices by the circulation
of the superconducting velocity. We revealed several unique
states combining vortices and skyrmions, their possible
reconfiguration with varied magnetic field, as well as the novel
temporal and field-induced transitions between vortical and skyrmionic states.
\end{abstract}


\maketitle

\section{Introduction}

Strontium ruthenate, Sr$_2$RuO$_4$, is according to theoretical predictions 
the best candidate to date to host $p\,$-wave superconductivity. 
Generally speaking, the order parameter 
in superconductors describes the spatial profile of the gap function, $\Delta_{ij}(k)$. 
The order parameter in $p\,$-wave superconductivity is an odd function of the wave vector ${\textbf k}$, 
unlike the s-wave superconductors where it is an even function of ${\textbf k}$. \cite{Maeno} 
Following the notation of Balian and Werthamer, the $p\,$-wave order parameter reads \cite{Rice, Sigrist}

\begin{equation}
\hat{\Delta}(k)=\left [
 \begin{array}{cc}
  -d_x(k)+i d_y(k) & d_z(k) \\
  d_z(k) & d_x(k)+i d_y(k)
 \end{array}
 \right ],
\end{equation}

\noindent or in a short notation $\hat{\Delta}(k)= i\, (\vec{d}(k)\cdot\hat{\sigma})\,\sigma_y$,
where $\vec{d}(k)$ transforms as a vector under rotations and $\sigma_i$ are Pauli matrices.
Microscopic calculation of the superconducting gap is a highly demanding task
that requires a detailed knowledge of the pairing mechanism which in many 
cases is not available. What remains then is to exploit all the symmetries
(continuous and discrete) exhibited by the material under consideration and 
build a model that will depend on certain number of parameters.   
The possible superconducting order parameters that have been reported 
for $p\,$-wave superconductors required a detailed description of 
the crystal structure of the considered material. \cite{Rice, Sigrist} In that respect, 
strontium ruthenate (SRO) is a layered perovskite with a crystal structure 
similar to the well known high-T$_c$ superconductor (La,Sr)$_2$CuO$_4$,
where oxygen ions at the corners of an octahedron surround the body-centered 
Ru ion. \cite{Maeno, Mackenzie} The planar layers of RuO$_2$ are separated by 
Sr layers that stack along the highly symmetric axis $c$. The Fermi surface of 
strontium ruthenate contains three sheets arising from the binding of the Ru and O ions 
within the same layer. Bindings between the RuO$_2$ layers are weak due to the long separation 
of the interplanar RuO$_6$ octahedra. The Fermi sheets $\alpha$ and $\beta$ are both 
one dimensional (1D), while the $\gamma$ sheet is two dimensional (2D).
For sufficiently high fields along the $c$ direction, the Zeeman splitting
between electrons with opposite spins demands vector $\vec{d}(k)$ to lie in the RuO$_2$
layer or the basal plane, \cite{Rice, SkChung} i.e $d_z(k)=0$.  Thus, 
just two components of vector $\vec{d}(k)$ are left unknown. 
A rigorous analysis beyond the scope of this work found that among
the possible representations for $d_x\pm id_y$, there is one which is 2D and 
consequently breaking time-reversal symmetry. \cite{Rice, Sigrist} 
This representation is suitable for strontium ruthenate since muon-spin relaxation 
($\mu$SR) measurements detected spontaneous magnetization below its superconducting 
critical temperature, $T_c = 1.5$ K, \cite{Luke, Mackenzie} fitting well to what was predicted 
in a formalism where the superconducting order parameter has at least two components. 
This 2D representation introduces an internal degree of freedom, the chirality, 
that has been widely discussed and reported in recent works. 
\cite{das_sarma, ivanov, halfQV, Read-Green} For example, the work of Chung \emph{et al.} 
provided the parameters needed to stabilize a vortex which is not coupled
to the electromagnetic potential and which carries half of the vorticity of 
a conventional vortex -in other words, a half quantum vortex (HQV). Skyrmion states carrying 
topological charge defined by the Hopf invariant, \cite{expskyone, Garaud} are 
another hallmark of chiral superconductivity, and were obtained as stable
solutions in bulk $p\,$-wave superconductors within the GL model. \cite{Garaud}  

In this work we report various full vortex (FV), HQV and skyrmion
states as the fundamental solutions of the GL model for \emph{mesoscopic} 
chiral $p\,$-wave superconductivity. \cite{Sigrist, SkChung}
It is well known in conventional s-wave superconductivity that confinement 
can stabilize superconducting configurations which in bulk systems
are energetically unfavorable or even unattainable, e.g. non-Abrikosov vortex 
lattices, or vortices with phase winding $\phi=2\pi n$, with $n>1$ (giant vortices).
\cite{gvs} Similarly, in mesoscopic chiral 
$p\,$-wave superconductors HQVs have been predicted to exist, owing the
reduction of their otherwise divergent energy to the low dimensionality of 
the system. \cite{SkChung} We found these HQV states in multiple forms, 
but also FV and skyrmionic states and transitions between them 
as a function of the external magnetic field applied perpendicularly to the
sample. We employed the time-dependent theoretical formalism, which allowed us
to observe novel temporal transitions as well, related to peculiar entry
and arrangement of HQVs and their temporal transformations into other topologies.

The paper is organized as follows. Sec. \ref{analytical} presents the theoretical
formalism and our analytical analysis of the first GL equation and the
superconducting current. The boundary conditions imposed on our equations
are derived from the latter expression. Sec. \ref{Results} then
summons our findings for the superconducting configurations composed of
HQV, FV and skyrmion states, obtained at weak coupling and 
considering a cylindrical Fermi surface. The transitions between states
of interest as a function of the magnetic field are discussed 
in Sec. \ref{transitions}, while the temporal transformations are shown 
in Sec. \ref{time_evol}. The effect of anisotropy on the topological, 
vortical and skyrmionic entities is analyzed in Sec. \ref{aniso}.
Our findings ans conclusions are summarized in Sec. \ref{summary}.

\section{\label{analytical}Theoretical Formalism}

After the above brief general description of strontium ruthenate, 
in what follows we show the Ginzburg-Landau (GL) equations that the order parameter, 
$\vec{\Psi}=(\psi_x,\psi_y)^T$ must satisfy. The order parameter has two components 
(is chiral) as a consequence of the 2 dimensional representation ($\Gamma_5^{\pm}$) 
of the tetragonal group $D_{4h}$. \cite{Sigrist} The expansion of the GL free energy density 
up to fourth order in $\psi_{x,y}$, that fulfills the group symmetries, reads

\begin{eqnarray}
\label{freecomp}
\mathscr{F} &=&  K \left ( |D_x \psi_x |^2+|D_y \psi_y|^2\right ) +  k_1\left( |D_x \psi_y |^2 +|D_y \psi_x |^2\right) \nonumber \\
 & + & 2\, {\rm Re} \Bigl\{k_2 D_x\psi_x (D_y\psi_y)^* + k_3 D_x \psi_y (D_y \psi_x )^*\Bigr\} \!-\! \alpha\,|\vec{\Psi}|^2 \nonumber \\
 & + & \beta_1|\vec{\Psi}|^4 + \beta_2(\psi_x^*\psi_y - \psi_x\psi_y^*)^2 + \beta_3|\psi_x|^2|\psi_y|^2\, ,
 \end{eqnarray}

\noindent where $\alpha$, $k_i$ and $\beta_i$, with $i=1,2,3$, are parameters
that depend on the details of the Fermi surface of the material under consideration. $K=\sum_i k_i$, and $D_{x,y}$
denote the components of the covariant derivative. The time-dependent Ginzburg-Landau (TDGL)
equations, used in our numerical approach, \cite{MiloGeurts} are the set of coupled differential equations 
for the superconducting order parameter, $\vec{\Psi}$, and the vector potential $\vec{A}$, \cite{Gropp}

\begin{eqnarray}
 \frac{\hbar^2}{2m_sD}\Big(\frac{\partial}{\partial t}+\frac{2ie}{\hbar}\varphi\Bigl)\vec{\Psi}&=&
-\frac{\delta\mathscr{F}}{\delta\vec{\Psi^*}}, \\
 \frac{\sigma}{c}\Big(\frac{1}{c}\frac{\partial\vec{A}}{\partial t}+\vec{\nabla}\varphi\Bigl)&=&
-\frac{\delta\mathscr{F}}{\delta\vec{A}}-\frac{1}{4\pi}\vec{\nabla}\times\vec{B}\, ,
\label{TDGL_Eq}
\end{eqnarray}

\noindent where $\varphi$ is the scalar electric potential, $\vec{B}$ is the magnetic induction, 
$m_s$ is the effective mass, $D$ is the phenomenological diffusion coefficient, 
and $\sigma$ the electrical conductivity. For convenience we set $\hbar=1$ and $m_s=1/2$. 
The second GL equation [Eq. (\ref{TDGL_Eq})] is discarded in this work since the diamagnetic effects 
of superconductors are vanishingly small for a thin (effectively 2D) mesoscopic geometry.
We use the symmetric gauge for the vector potential, $\vec{A}=( \vec{r}\times\vec{H} )/2$, 
with the magnetic field ($\vec{H}$) directed along $\hat{z}$. 
The scalar electric potential is set to zero since neither charges nor 
external currents are considered in this work. In dimensionless units, where distance is 
scaled to the coherence length, $\xi=\sqrt{\frac{1}{\alpha}}$,  
time to $t_0=\frac{\xi^2}{D}$, magnetic field to the upper bulk critical field
$H_{c2}=\frac{c}{2|e|\xi^2}$, and the superconducting order parameter 
to $\Delta_+=\sqrt{\frac{\alpha}{2\beta_1}}$, the first TDGL equation becomes
\begin{widetext}
\begin{eqnarray}
 \frac{\partial \vec{\Psi}}{\partial t} &\!=\!&
\left[
 \begin{array}{cc}
  \frac{K+k_1}{2} \vec{D}^2 + \frac{k_2-k_3}{2i}[D_x,D_y] & (k_2+k_3)\,\Pi_+^2 \\
  (k_2+k_3)\,\Pi_-^2 & \frac{K+k_1}{2} \vec{D}^2 - \frac{k_2-k_3}{2i}[D_x,D_y]
 \end{array}
\right]
\left(
\begin{array}{c}
\psi_+ \\
\psi_-
\end{array}
\right) 
+ \vec{\Psi}\Bigl(1-\frac{1+\tau}{2}|\vec{\Psi}|^2 \pm\frac{\tau}{2}\vec{\Psi}^*\hat{\sigma}_z\vec{\Psi}\,\Bigl) \, ,
 \label{GL_gnral}  
 \end{eqnarray}
 \end{widetext}
\noindent where $\Pi_\pm = \frac{1}{\sqrt{2}}(D_x \pm i D_y)$, $\psi_\pm=\psi_x \pm i \psi_y$ and
$\tau=\beta_2/\beta_1$. A straightforward calculation reveals the following important result,
$[D_x,D_y]=iH$, which leads the operators $\Pi_\pm$ to satisfy the commutator: $[\Pi_+,\Pi_-]=H$. 
The external magnetic field, being constant, can be factored out 
from the above commutators, leading to $[\tilde{\Pi}_+,\tilde{\Pi}_-]=1$, 
which defines the algebra behind the Landau levels; $\tilde{\Pi}_\pm=\Pi_\pm/\sqrt{H}$.
This algebra is defined through the following commutators: $[\hat{N},\tilde{\Pi}_+] = -\tilde{\Pi}_+$,
$[\hat{N},\tilde{\Pi}_-] = \tilde{\Pi}_-$; where $\hat{N}=\tilde{\Pi}_+\tilde{\Pi}_-$ is the 
particle number operator. Within the weak-coupling limit and considering a cylindrical Fermi surface 
($\gamma$ sheet), all the $k_i$ parameters are equal to 
$\langle v_x^2v_y^2\rangle / \langle v_x^4 \rangle = 1/3$, where brackets $\langle\,\rangle$ 
denote averaging over the Fermi surface. \cite{Agterberg} For this case the first GL equation reads
\begin{equation}
 \partial_t\vec{\Psi} = \frac{2}{3}\Bigl[ \vec{D}^2 + \Pi_+^2\hat{\sigma}_+ + \Pi_-^2\hat{\sigma}_- \Bigr]\vec{\Psi} 
 + \vec{\Psi}\Bigl(1-\frac{3|\vec{\Psi}|^2}{4} \pm\frac{\vec{\Psi}^*\hat{\sigma}_z\vec{\Psi}}{4}\Bigl),
 \label{GL_cyl}
 \end{equation}
\noindent where $\hat{\sigma}_\pm = ( \hat{\sigma}_x \pm i \hat{\sigma}_y )/2$, are 
pseudospin or chiral operators acting on the space span by $\psi_\pm$. 
Ignoring the nonlinear terms (linearized case), 
it is straightforward to show that the superconducting order
parameter must be of the form: $\vec{\Psi} = (\phi_N, \phi_{N-2})^T$, 
where $\phi_N$ is the state corresponding to the Landau level $N$. \cite{Agterberg, Furusaki, Bor-Luen, Zhu}
Within the superconducting formalism the number $N$ turns out to be the vorticity of the order parameter.
Then, one concludes that for chiral $p\,$-wave superconductors there is a vorticity difference
two between the components of the superconducting order parameter. The full GL equations, i.e. 
the linearized equation plus the nonlinear terms, are a complicated set of partial 
differential equations with unknown analytical solutions. Therefore, in this work we
solve this problem numerically. Due to the mesoscopic dimension of the sample under consideration, 
proper boundary conditions must be incorporated in the GL equations in order to pose the problem well. 
In what follows, the superconducting current is calculated for the general case, which includes the 
specific case where all $k_i$'s are equal to 1/3, and from this expression the boundary 
conditions for the first GL equation are derived. The superconducting current density, 
defined as the negative functional derivative of the GL free energy density with respect 
to the vector potential, for chiral $p\,$-wave superconductors is
\begin{widetext}
\begin{equation}
 \vec{J} = {\rm Im}\biggl\{\frac{K\!+\!k_1}{4}\Bigl(\psi_+^*\vec{D}\psi_+ + \psi_-^*\vec{D}\psi_- \Bigr) 
+ \frac{k_2+k_3}{2\sqrt{2}}\Bigl (\vec{\Psi}^*\Bigl[\Pi_+\hat{\sigma}_+ + \Pi_-\hat{\sigma}_-\Bigr]\vec{\Psi}\,\hat{\imath} 
+ i\,\vec{\Psi}^*\Bigl[\Pi_+\hat{\sigma}_+ - \Pi_-\hat{\sigma}_-\Bigr]\vec{\Psi}\,\hat{\jmath} \Bigr) 
+ \frac{k_2-k_3}{4} \vec{\Psi}^* \hat{S}_y\vec{D}\hat{\sigma}_z \vec{\Psi}\biggr\},  
\label{sucurr_gnral}
\end{equation}
\end{widetext}

\noindent where $\hat{\imath}$, $\hat{\jmath}$ form the canonical base in Cartesian coordinates. 
The set of operators ($\hat{\sigma}_\pm$ and $\hat{\sigma}_z$) act on $\psi_\pm$, while
$\hat{S}_y$ acts on $\{\hat{\imath}, \hat{\jmath}\}$. The superconducting current contains 
mainly three contributions defined by the following factors, $(K+k_1)/4$, $(k_2+k_3)/2\sqrt{2}$,
and $(k_2-k_3)/4$. The first one arises from the conventional term $\vec{D}^2$ in Eq. (\ref{GL_gnral}), 
the second one (we name chiral) is due to the internal degree of freedom (chirality)
that appears in Eq. (\ref{GL_gnral}) in the form of two nondiagonal terms. Finally, the third  
contribution arises from the diagonal terms ($\pm[D_x,D_y]$) in Eq. (\ref{GL_gnral}), 
and resembles the Zeeman interaction between the magnetic field and the electron spin. 
In this case $k_2-k_3$ plays the role of the magnetic moment. 
The boundary conditions imposed on Eq. (\ref{GL_gnral}) for our square
mesoscopic sample are given as:
\begin{eqnarray}
&\left .\
\begin{array}{r}
\psi_+ - \psi_- = 0 \\
D_y\psi_+ + D_y\psi_- = 0 
\end{array}
\right\}
\quad \text{at north and south sides}, 
\nonumber \\
&\left.\
\begin{array}{r}
\psi_+ + \psi_- = 0 \\
D_x\psi_+ - D_x\psi_- = 0 
\end{array}
\right\}
\quad \text{at east and west sides.}
\label{bcond}
 \end{eqnarray}
It is straightforward to show that the boundary conditions of Eq. (\ref{bcond}) set 
the perpendicular current  at the edges to zero, i.e. they impose specular reflection in the 
chiral $p\,$-wave superconductor. \cite{Sigrist, Furusaki, Bor-Luen} 
It is important to remark also that they are parameter-independent, so they provide 
the proper boundary conditions for Eq. (\ref{GL_cyl}) but also for the most general case
of Eq. (\ref{GL_gnral}). With Eq. (\ref{bcond}) we have completed the set of equations needed
for the GL description of a chiral $p\,$-wave 
mesoscopic superconductor. Eq. (\ref{GL_gnral}) is numerically solved using 
finite differences and the link variables technique of Ref. [\onlinecite{Gropp}] 
on a square lattice with mesh grid $h_x=h_y=0.1$. On the other hand,
the temporal derivative is discretized using the Runge-Kutta 
method of first order. Before concluding this section, 
we give the reduced expression for the dimensionless free energy, 
since it allows us to find not only the lowest energy (ground) states 
but also the stable states with slightly higher energies (metastable states) 
The free energy reads: 
\begin{equation}
 \frac{F}{F_0} = \frac{1}{2}\int\!dV \Bigl\{ (1+\tau)|\vec{\Psi}|^4  + \tau(\vec{\Psi}^*\hat{\sigma}_z\vec{\Psi})^2 \Bigr \},
\end{equation}
\noindent where $F_0=\Delta_+^2/\xi^2$ is the bulk free energy at zero field.

\section{\label{Results}Isotropic Case (Cylindrical Fermi surface)}
%\subsection{}
%\subsubsection{}

\begin{figure}[htb]
\center
\includegraphics[clip=true,trim={0.1cm 0.2cm 0.3cm 0.0cm}, scale=1.3]{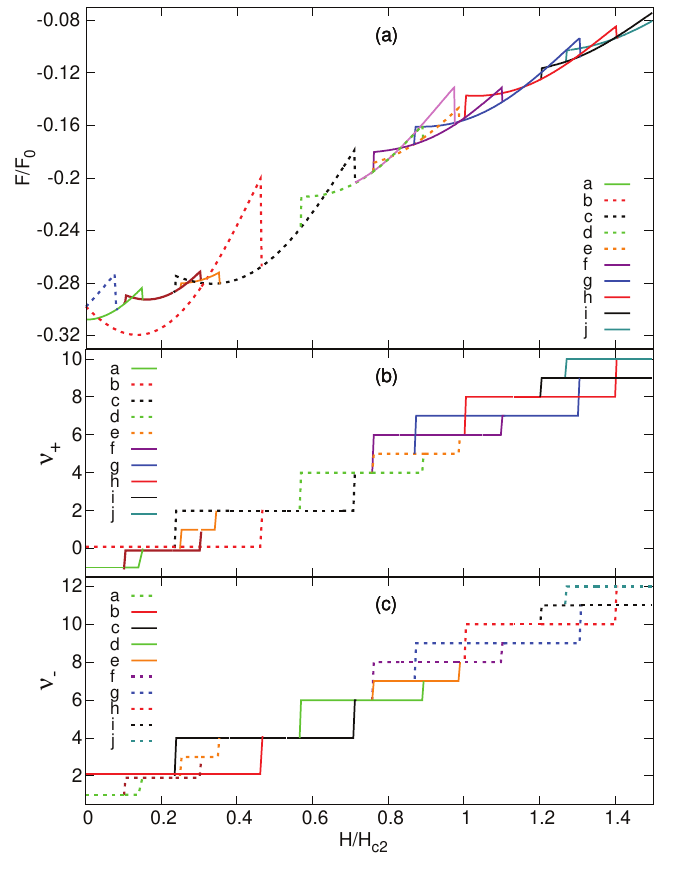}
\caption{(Color online) (a) Free energy in units of the bulk condensation energy at zero field ($F_0$) 
as a function of the external magnetic field in units of the bulk upper critical field ($H_{c2}$), 
for a square mesoscopic sample of size 8$\xi\times$8$\xi$. Letter labels denote different
found ground states. Some metastable states (not labeled) are also shown in this figure. 
Vorticity of components $\psi_+$ and $\psi_-$ of the ground states of panel (a)
are shown in (b) and (c) respectively. The difference in vorticity ($\nu_+\!-\!\nu_-\!=\!2$) 
between the components is in perfect agreement with the analytically predicted solution 
$\vec{\Psi}=(\phi_N,\phi_{N-2})^T$.}
\label{free_dk00}
\end{figure}

\begin{figure}[htb]
\includegraphics[trim={0.0cm 0.0cm 0.1cm 0.0cm}, clip=true, scale=1.3]{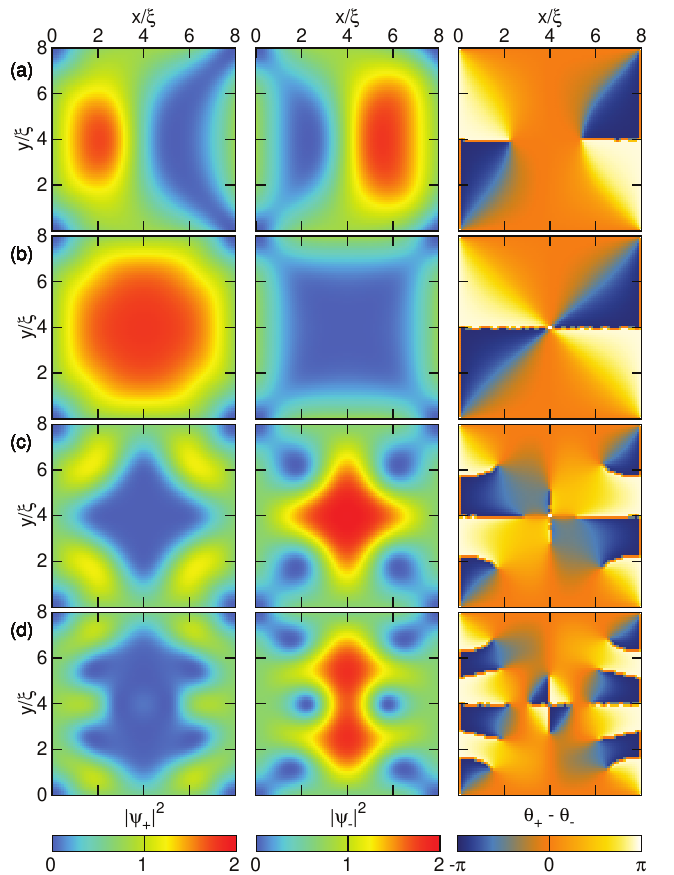}
\caption{(Color online) Ground states \emph{a - d} of Fig. \ref{free_dk00}. Left and central columns show the contour 
plots of the superconducting densities components $|\psi_+|^2$ and $|\psi_-|^2$, respectively. 
Right column shows the difference between the angular phases of the components, 
i.e. $\theta_+ - \theta_-$.}
\label{st_dk00_pm}
\end{figure}

The results obtained using Eq. (\ref{GL_cyl}) for a square 8$\xi\times$8$\xi$ sample 
are summarized in Fig. \ref{free_dk00}, showing the dimensionless 
free energy and the vorticity of vector $\vec{\Psi}$ as a function of the external 
magnetic field $H$. Panels (b) and (c) show the vorticity of the ground states
of our superconducting sample, labeled \emph{a-j} in Fig. \ref{free_dk00} (a), 
where $\nu_{+(-)}$ is the vorticity of component $\psi_{+(-)}$. 
Note that both $\nu_+$ and $\nu_-$ remain constant along the stability curves of 
each state in panel (a), and as such are good identification numbers for these states.  
Contour plots in Fig. \ref{st_dk00_pm} show the order parameter $\vec{\Psi}$ corresponding
to the ground states \emph{a - d}. While the left and central columns of Fig. \ref{st_dk00_pm} 
show contour plots of the superconducting density of each component, $|\psi_+|^2$ and $|\psi_-|^2$, 
respectively, the third column shows the difference between the angular phases of the 
components, i.e. $\theta_+-\theta_-$. 

The ground state \emph{a} of Fig. \ref{st_dk00_pm} shows one anisotropic vortex in each 
component, i.e. vorticity  $\nu_+=-1$ in component $\psi_+$ and 
$\nu_-=1$ in component $\psi_-$. The contour plots of the ground state \emph{b} in Fig. \ref{st_dk00_pm}, 
show the vortex free state in component $\psi_+$ and the giant vortex \cite{gvs} 
with vorticity $\nu_-=2$ in component $\psi_-$. The subsequent ground state \emph{c} has
vorticity $\nu_+=2$ and $\nu_-=4$, where $|\psi_-|^2$ contains four vortices close to the corners, 
meanwhile $|\psi_+|^2$ shows a pronounced depletion around the center of the sample. 
The corresponding phase difference figure reveals that the depletion in component $\psi_+$
is a consequence of two vortices and two vortex-antivortex pairs there. The ground state \emph{d}
has six vortices in $|\psi_-|^2$ in full agreement with the vorticity reported in Figs. \ref{free_dk00}
(b) and (c) ($\nu_+=4$ and $\nu_-=6$). However, the density $|\psi_+|^2$ fails to convincingly show
any signature of a vortex. The vorticity $\nu_+=4$ of component $\psi_+$ is visible in 
the phase difference figure 2 (d), where 10 discontinuities are found along the 
edges as a consequence of six vortices from $\psi_-$ and 
four from $\psi_+$. Four vortex-antivortex pairs at the center of the sample are also visible  
in this contour plot, but do not affect the total vorticity.
%
%\begin{figure}[htb]
%\center
%\includegraphics[clip=true,trim={0.0cm 0.0cm 0.0cm 0.0cm}, scale=0.7]{dk00_vlow}
%\caption{Vorticity of the ground states of Fig. \ref{free_dk00}.
%(b) and (c) panels show the vorticity of components $\psi_+$ and $\psi_-$, respectively.
%The difference in vorticity ($\nu_+\!-\!\nu_-\!=\!2$) between the components is in perfect agreement with 
%the theoretically predicted solution $\vec{\Psi}=(\phi_N,\phi_{N-2})^T$.}
%\label{vor_dk00}
%\end{figure}

\begin{figure}[htb]
\includegraphics[trim={0.35cm 0.0cm 0.0cm 0.1cm}, clip=true, scale=1.7]{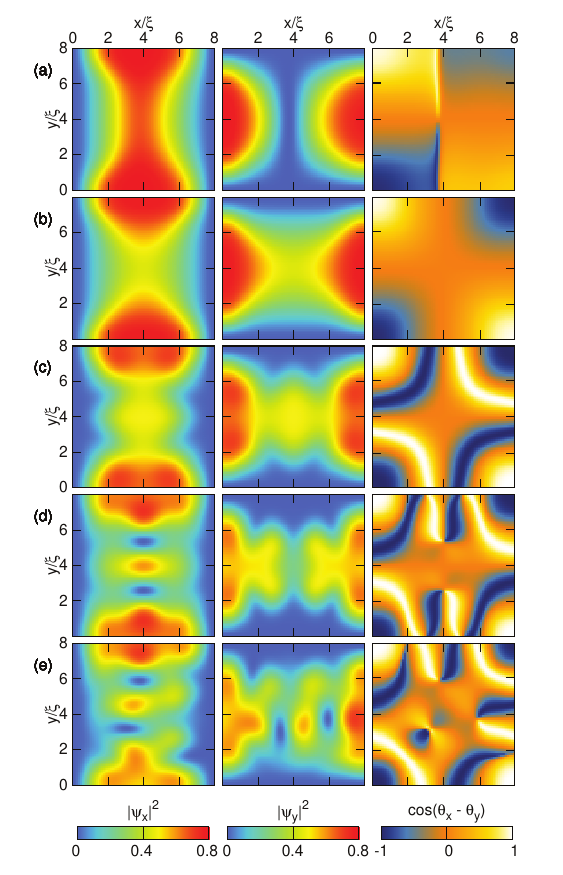}
\caption{(Color online) Ground states \emph{a - e} of Fig. \ref{free_dk00}, plotted correspondingly 
in panels (a) - (e). Left and central columns show the contour plots
of the superconducting densities components $|\psi_x|^2$ and $|\psi_y|^2$, respectively. 
Right column shows $\cos{(\theta_x-\theta_y)}$, where $\theta_{x,y}$ are the 
angular phases of components $\psi_x$ and $\psi_y$.}
\label{st_dk00_low}
\end{figure}

\begin{figure}[htb]
\includegraphics[trim={0.35cm 0.0cm 0.0cm 0.1cm}, clip=true, scale=1.7]{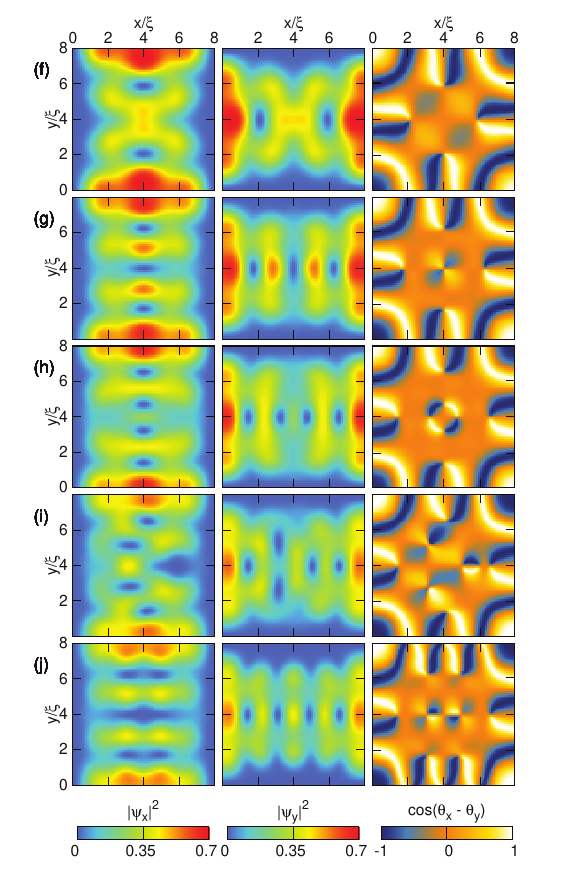}
\caption{(Color online) Ground states \emph{f - j} of Fig. \ref{free_dk00}, plotted correspondingly in
panels (f) - (j). Displayed quantities are the same as in Fig. \ref{st_dk00_low}.}
\label{st_dk00_high}
\end{figure}

From the comparison between Figs. \ref{st_dk00_pm} (c) and (d) one sees that with increasing the magnetic 
field the component $\psi_-$ dominates its partner component $\psi_+$. The dominance of 
$\psi_-$ over $\psi_+$, especially at high fields impedes the proper description of the vortex 
configuration in the latter component. In order to describe the components of 
the order parameter on an equal footing, a more suitable representation is
in terms of $\psi_x$ and $\psi_y$. Figs. \ref{st_dk00_low} and \ref{st_dk00_high}
show contour plots of $|\psi_x|^2$, $|\psi_y|^2$ and $\cos(\theta_x-\theta_y)$ for 
ground states \emph{a - j} of Fig. \ref{free_dk00}. Fig. \ref{st_dk00_low} (a) shows 
$\cos(\theta_x-\theta_y)$ for ground state \emph{a}
(from now on called the phase difference figure), and reveals a linear domain wall. Its extension 
across the sample coincides with the stripe where density $|\psi_y|^2$ vanishes.
On the other hand, the partner component, $\psi_x$, is free of vortices. 
Ground states \emph{b} and \emph{c} look similar in both densities, although from the comparison between 
their phase difference figures in Fig. \ref{st_dk00_low} (b) and (c) respectively, 
we see four domain walls in state \emph{c} and none in state \emph{b}. 
The domain walls (DWs) of ground state \emph{c} define a path where the 
difference between the angular phases of components $\psi_x$ and $\psi_y$ are $0$ or
$\pi$, i.e. $\theta_x-\theta_y=0, \pi$.   
Ground state \emph{d} shows two vortices in density $|\psi_x|^2$ and none in $|\psi_y|^2$, 
while its corresponding phase difference figure shows the four domain walls of ground state \emph{c}
plus two other alternating domain walls that weakly connect the former ones.
The contour plots of Fig. \ref{st_dk00_low} (e), for state \emph{e} show clearly two vortices in each component.
They look indistinguishable just from the analysis of their densities, but 
their phase difference figure reveals that there are two vortices, one in each component, 
that combine to produce a different signature from the remaining vortices. While the  
uncorrelated vortices lead to the formation of the alternating domain walls towards sample edges,
the pair of correlated vortices align their cores and do not show any domain wall between them. 
The alternating domain wall is therefore the signature of the half quantum vortex (HQV), 
in contrast to the other signature without domain wall that corresponds to the full vortex (FV).

Fig. \ref{st_dk00_high} shows the remaining ground states \emph{f - j} of Fig. \ref{free_dk00}.
Both densities in ground state \emph{f} clearly show two vortices in each component, 
which are indeed four HQVs according to the corresponding phase difference figure.   
Ground states  \emph{g} and \emph{h} show one common feature,  having different number of HQVs per 
component, but all of them aligned vertically in component $\psi_x$ and horizontally in 
component $\psi_y$. On the other hand, the corresponding phase difference figures 
for states \emph{g} and \emph{h} show that: (i) two vortices, one per component, combine to form one FV
in state \emph{g}, and (ii) four vortices, two per each component, combine to form one skyrmion in state \emph{h}. 
The signature of the skyrmion is shown here for the first time: four alternating domain walls which are connected 
into a circular structure. \cite{Garaudprl, Garaudech} The skyrmion state here of course differs from those of 
magnetic materials due in physics and the formation mechanism. \cite{expskyone,expskytwo,expskythree}
Nevertheless, their topological properties remain similar, as will be presented later.
The phase difference figure of the ground state \emph{i} shows
four DWs around the corners, four HQVs close to the edges and three FVs
in the center. What draws attention in all three contour plots of Fig. \ref{st_dk00_high} (i) 
is that there are five vortices in each component (fractional vortices), and among them three align
their cores to form FVs according to the corresponding phase difference figure. The triangular array
formed by them resembles the consequences of vortex-vortex repulsion in conventional type II superconductors. 
Therefore, this supports our initial premise that the FV in our analysis is the usual 
Abrikosov vortex of conventional superconductivity. Finally, the phase difference figure of 
the ground state \emph{j} shows four DWs, six HQVs and two FVs. 
One systematic comparison of the phase difference figure
of ground states \emph{f - j} clearly shows that HQV and FV are indeed very different states. 
While FVs are formed in the sample center, being favored by confinement, 
all the HQVs remain close to the sample edges. In order to explain this difference 
the following subsection discusses the calculated superconducting currents
in the sample.

\begin{figure}[htb]
\centering
\includegraphics[clip=true,trim={1.1cm 0.5cm 7cm 13.5cm}, scale=0.6]{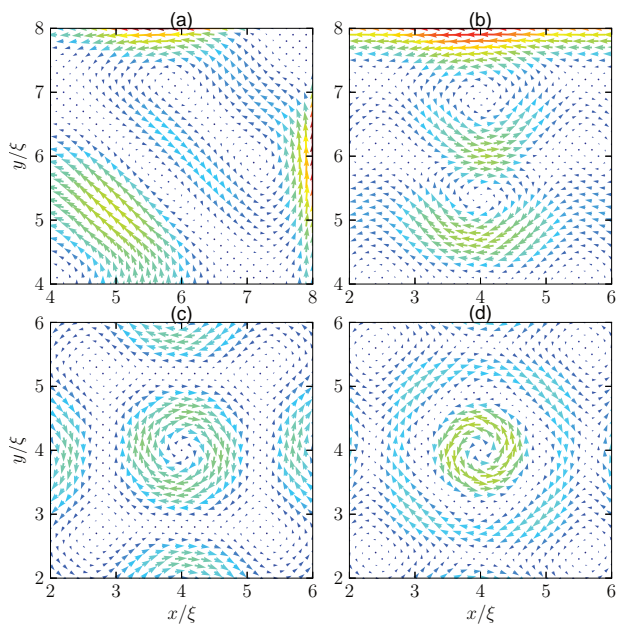}
\caption{(Color online) Superconducting currents around: (a) the upper right DW of Fig. \ref{st_dk00_low}(c), 
(b) the upper HQV of Fig. \ref{st_dk00_low}(d), (c) the FV of Fig. \ref{st_dk00_high}(g), 
and (d) the skyrmion of Fig. \ref{st_dk00_high}(h).}
\label{curr_four}
\end{figure}
\begin{figure}[htb]
 \includegraphics[clip=true, trim={0.2cm 0.0cm 0.7cm 0.0cm}, scale=0.7]{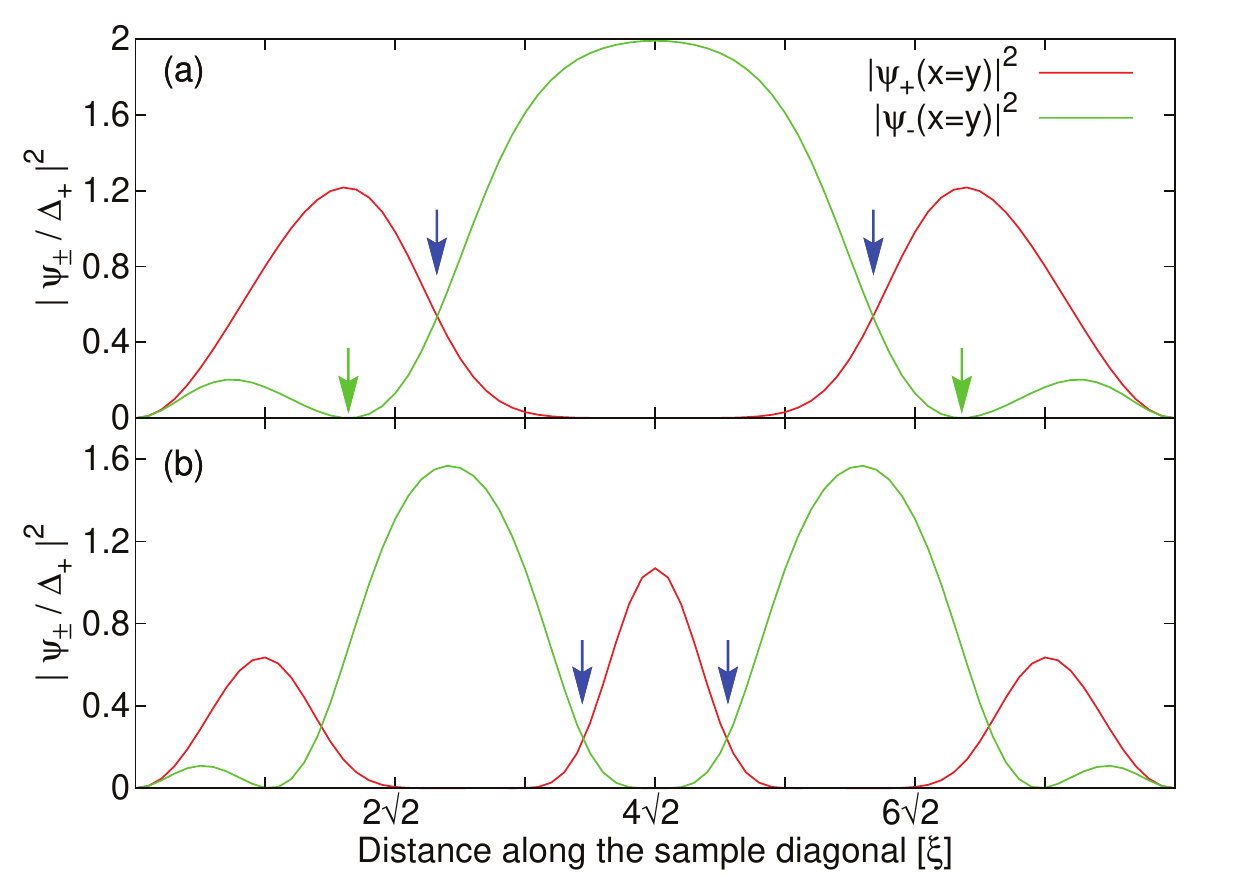}
 \caption{(Color online) Diagonal profiles of the contour plots $|\psi_\pm|^2$ corresponding to ground states
 \emph{c} and \emph{h} of Fig. \ref{free_dk00}, shown in panels (a) and (b), respectively. 
 Blue and green arrows indicate the DW and vortex core locations, respectively.}
 \label{DW_profile}
\end{figure}
\begin{figure}[htb]
\centering
\includegraphics[clip=true,trim={1.52cm 0.5cm 4cm 13.5cm}, scale=0.54]{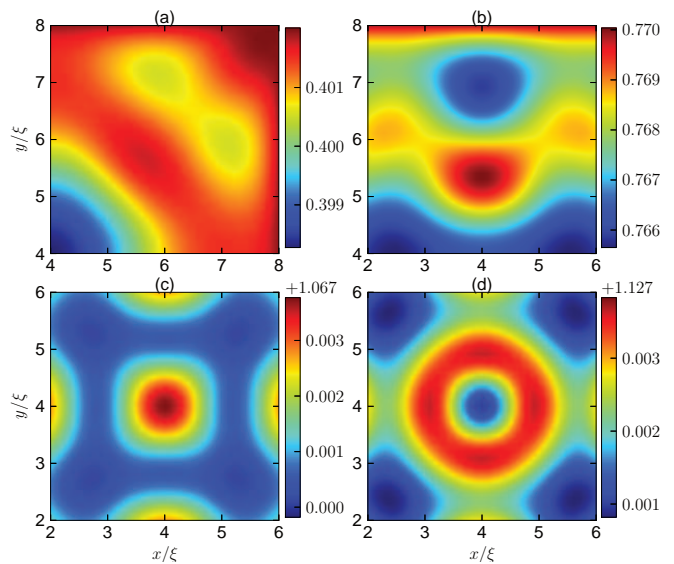}
\caption{(Color online) Contour plots of the magnetic induction corresponding to: (a) the 
supercurrents of the DW of Fig. \ref{curr_four}(a), (b) the HQV of Fig. \ref{curr_four}(b),
the FV of Fig. \ref{curr_four}(c), and (d) the skyrmion of Fig. \ref{curr_four}(d).}
\label{indmag_four}
\end{figure}

\begin{figure*}[htb]
\center
\includegraphics[clip=true,trim={0.75cm 4.0cm 2.0cm 3.0cm}, scale=0.6]{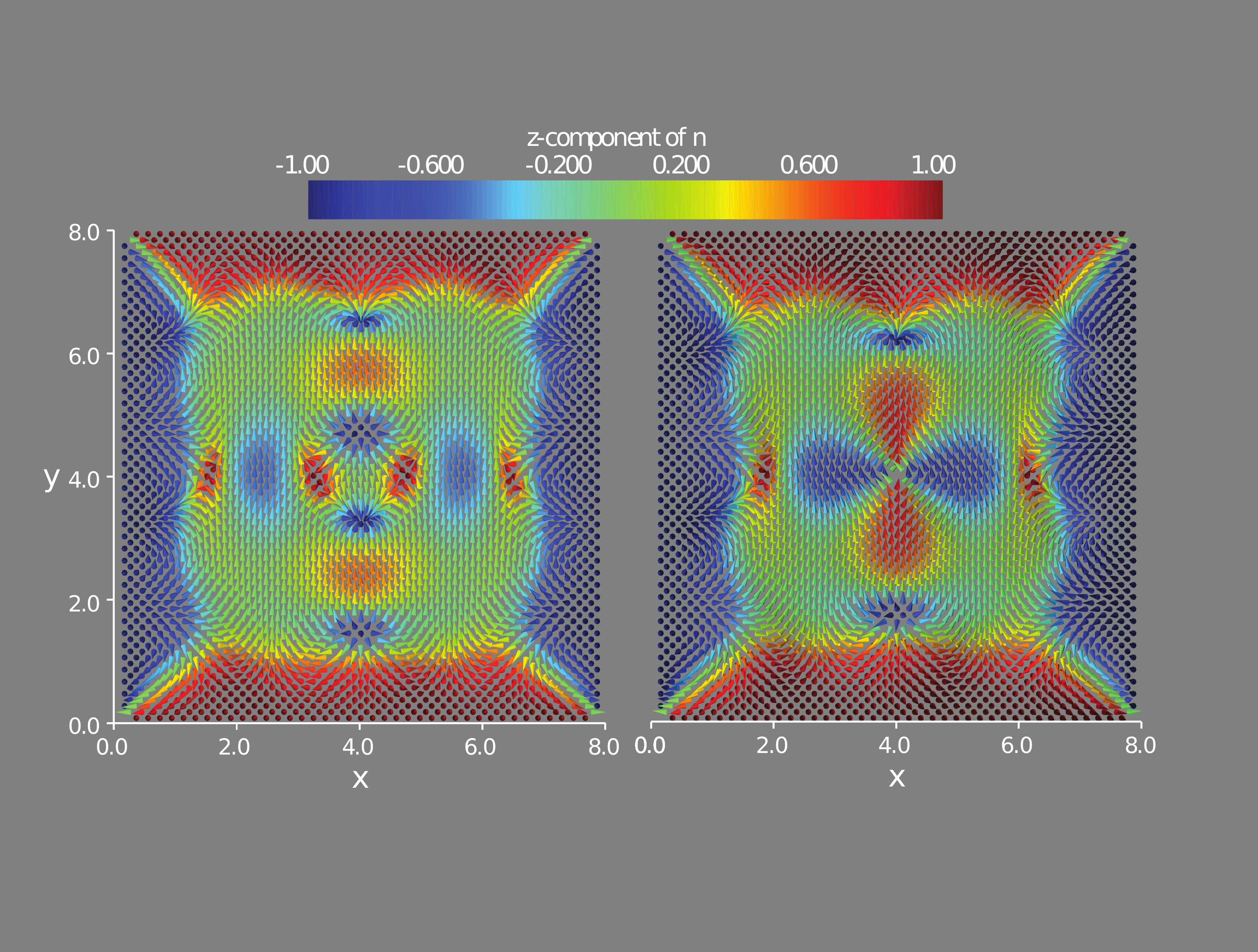}
\caption{(Color online) Textures of the ground states \emph{h} and \emph{g} of Fig. \ref{st_dk00_high},
according to the mapping 
$\hat{n}=\vec{\Psi}^\dagger\hat{\sigma}\vec{\Psi}/\vec{\Psi}^\dagger\!\cdot\vec{\Psi}$,
where $\hat{\sigma}$ are the Pauli matrices. Colors show the amplitude of the z-component
of $\hat{n}$.}
\label{sky_field}
\end{figure*}
So far, DWs, HQVs, FVs and skyrmions have been distinguished in this work 
according to their signatures in the phase difference plots. 
The superconducting current, the physical quantity intertwined with the magnetic field, 
also allows us to identify more characteristic features of the novel topological solutions.  
Figs. \ref{curr_four} (a)-(d) show the supercurrents around one DW, HQV, FV and 
skyrmion, respectively. Fig. \ref{curr_four} (a) zooms in the supercurrents around
the right-top DW of Fig. \ref{st_dk00_low}(c). Its supercurrents are flowing from the upper left
to the bottom right corner, where the DW outer and inner screening currents flow counter-clockwise. 
Aside the DW current, there are two narrow stripes where the supercurrent vanishes, 
consequence of the continuity of the supercurrent field and the compensation between
the screening and the DW currents. In order to understand better the origin of the DW currents,
Fig. \ref{DW_profile}(a) shows the line profiles of the corresponding superconducting densities 
$|\psi_\pm|^2$ along the diagonal line defined by $y=x$. Light (green) arrows point towards  
the already seen vortex cores of component $\psi_-$ in Fig. \ref{st_dk00_pm}(c). Dark (blue) arrows
indicate the center of two DWs defined by the intersection where the densities $|\psi_-|^2$ and $|\psi_+|^2$
become equal. A straightforward calculation that exploits the previous definition: $|\psi_+|^2=|\psi_-|^2$,
leads to: $\psi_x\psi_y^*-\psi_y\psi_x^*=0$, or equivalently:  $\sin{(\theta_x-\theta_y)} = 0$,
in full agreement with the contour plots of the phase difference in Fig. \ref{st_dk00_low}(c),
where $\theta_x-\theta_y\! =\!0$, or $\theta_x-\theta_y\! =\!\pi$ along the DWs. 

The magnetic induction that corresponds to the DW supercurrents of Fig. \ref{curr_four}(a)
is shown in panel (a) of Fig. \ref{indmag_four}. It is calculated using the Maxwell equation,
\begin{equation}
 \tilde{\kappa\,}^2 \,\nabla\times\vec{B} =  \vec{J},
\label{maxwell}
 \end{equation}

\noindent where $\tilde{\kappa\,}^2 = \kappa^2/d$, with $\kappa=46$ being the GL 
parameter reported for SRO along the \emph{c} axis, \cite{Mackenzie} 
and $d$ being the sample thickness which we suitably choose 
to be $2\xi$. The contour plot of Fig. \ref{indmag_four}(a) shows that the magnetic induction corresponding 
to the DW is weak and strongly screened by the Meissner effect. This fact represents
an obstacle for the detection of DWs signatures in direct measurements of their
magnetic response such as in magnetic force microscopy (MFM) or scanning Hall probe 
microscopy (SHPM).

Fig. \ref{curr_four}(b) zooms in the supercurrents around the upper HQV 
of Fig. \ref{st_dk00_low}(d). It shows two adjacent counter-flowing streams
with the bottom one flowing clockwise and belonging to the HQV supercurrents, 
while the top one flows counter-clockwise and represents the screening currents.
The Meissner effect for the HQV is anisotropic due to the boundary conditions of Eq. (\ref{bcond}). 
From the supercurrent equation (\ref{sucurr_gnral}), and the local approximation
$\psi_y \approx 0$, or $\psi_+ \approx \psi_-$, drawn from Fig. \ref{st_dk00_low}(d), 
one easily obtains: $\vec{J} \approx Im\{ \psi_+^* D_x\psi_+ \hat{\imath} + \frac{1}{3}\psi_+D_y\psi_+\hat{\jmath}\} $.
After straightforward calculations and replacing the covariant derivative once
again one obtains:
$\vec{J} \approx |\psi_+|^2\bigl[ (\partial_x\theta \, \hat{\imath}+ \frac{1}{3}\partial_y\theta \, \hat{\jmath}) 
+ \frac{Hr}{2}(\sin{\phi} \, \hat{\imath} - \frac{1}{3}\cos{\phi} \, \hat{\jmath}) \bigr]$,
which draws attention since the screening currents are defining elliptical equipotential lines.  
Thus, the anisotropic screening of the superconductor towards the HQVs causes them 
to move along the easy-screening direction which in this case is along $\hat{y}$.
The contour plot of the magnetic induction corresponding to the supercurrents 
of Fig. \ref{curr_four}(b) is shown in Fig. \ref{indmag_four}(b). 
As expected from the two counter-flowing streams seen in the HQV supercurrents,
the magnetic induction also shows adjacent local maximum and local minimum.

Fig. \ref{curr_four}(c) zooms in the superconducting currents 
around the FV of Fig. \ref{st_dk00_high} (g), and shows that the FV currents flow clockwise and vanish as 
we move away from the FV core. This vanishing is due to the spatially isotropic Meissner effect,
unlike in a HQV, that screens the FV currents. As expected, its magnetic induction signature 
[see Fig. \ref{indmag_four}(c)] agrees well with that of the Abrikosov vortex. 

The superconducting currents around the skyrmion of
Fig. \ref{st_dk00_high} (h) are shown in Fig. \ref{curr_four} (d). Unlike the FV,
the skyrmion supercurrents clearly show outer and inner structures. The 
supercurrents of the outer structure flow clockwise while the supercurrents 
of the inner flow counter-clockwise. The skyrmionic DW of Fig. \ref{st_dk00_high}(h)
along with its supercurrents in Fig. \ref{curr_four}(d) shows cylindrical symmetry,
and one easily deduces that the same symmetry is present in densities $|\psi_\pm|^2$. 
Line profiles of $|\psi_\pm|^2$ then provide enough information to unveil the skyrmion supercurrents 
[see Fig. \ref{DW_profile} (b)]. The inner structure of the skyrmion is defined by: 
$\psi_-\!=\!0$ and $\vec{\nabla}\theta_+\!=\!0$, 
where $\theta_+$ is the angular phase of component $\psi_+$, which when replaced in 
the supercurrent of Eq. (\ref{sucurr_gnral}) yields 
$\vec{J}\approx \frac{K+k_1}{4}|\psi_+|^2\frac{Hr}{2}\hat{\phi}$, 
with $\hat{\phi}$ being the azimuthal vector in cylindrical coordinates. 
Hence, the supercurrents at the skyrmion core are mainly due to the Meissner effect. 
However, away from the skyrmion core the scenario changes since the circular DW of 
Fig. \ref{st_dk00_high} (h) is met, as indicated by arrows in Fig. \ref{DW_profile} (b). 
Close beyond the circular DW, we find that component $\psi_+$ drops to zero. 
Replacing in Eq. (\ref{sucurr_gnral}) $\psi_+ = 0$ and bearing in mind that one giant vortex is 
hosted in component $\psi_-$, the supercurrent in cylindrical coordinates becomes: 
$\vec{J}\approx \frac{K+k_1}{4}|\psi_-|^2(-\frac{2}{r} + \frac{H}{2}r))\hat{\phi}$.
The magnetic induction corresponding to the skyrmionic supercurrents of Fig. \ref{curr_four}(d)
is shown in the contour plot of Fig. \ref{indmag_four}(d). It clearly shows
one local minimum at the skyrmion core surrounded by one circular stripe of local maxima,
and as such can be directly imaged in magnetic measurements.
 
Finally, before concluding this section, we briefly describe the 
superconducting order parameter $\vec{\Psi}$ in terms of the 3-dim real 
vector field $\hat{n}$, defined by: \cite{Babaevrpc,Garaud}
\begin{equation}
 \hat{n} = \frac{\vec{\Psi}^\dagger\hat{\sigma}\,\vec{\Psi}}{\vec{\Psi}^\dagger\!\cdot\!\vec{\Psi}},
\label{sky_map}
 \end{equation}

\noindent which maps the complex spaces $\mathbb{C}\times\mathbb{C}$ of components
$\psi_x$, $\psi_y$ into the real space $\mathbb{R}^3$. A straightforward calculation 
yields: $\hat{n}=\sin{\alpha}\cos{\phi}, \sin{\alpha}\sin{\phi},\cos{\alpha}$, where
$\sin{\alpha}=\frac{2|\psi_x||\psi_y|}{|\psi_x|^2+|\psi_y|^2}$, 
$\cos{\alpha}=\frac{|\psi_x|^2-|\psi_y|^2}{|\psi_x|^2+|\psi_y|^2}$, and 
$\phi=\theta_y-\theta_x$. Then, the target space of mapping (\ref{sky_map}) is
the 2-dim sphere of radius one, $\mathbb{S}^2$. \cite{Babaevprl, Rosenstein} The topological invariant of the 
spaces that result from mapping $(\ref{sky_map})$ is defined by 
the integral \cite{Garaud, expskyone}
\begin{equation}
 \mathbb{Q} = \frac{1}{4\pi}\!\int\!\hat{n}\cdot(\partial_x\hat{n}\times\partial_y\hat{n})\,dx\,dy ,
\label{hopf_inv}
 \end{equation}

\noindent which is widely known as the Hopf invariant. One convenient interpretation of
this topological invariant is that it counts the number of times that the 3-dim real 
field ($\hat{n}$) wraps around the 2-dim sphere ($\mathbb{S}^2$). Left and right panels of 
Fig. \ref{sky_field} show the texture $\hat{n}$ for ground states \emph{h} and \emph{g} of 
Fig. \ref{st_dk00_high}, respectively. The texture for the skyrmion (left panel) differs from 
the texture for the FV (right panel) owing to the alternating circular DW characteristic of the 
former state. While at the skyrmion core, field $\hat{n}$ points towards $-\hat{\jmath}$, 
outside the skyrmion it points towards $\hat{\jmath}$. Along the DW that separates the skyrmion
core from the outside, the field texture whirls, therefore providing to the space 
the topological charge $\mathbb{Q}=-2$. The field texture that corresponds to the FV 
shows four lobes C4 symmetric profile where $\hat{n}$ changes smoothly. Unlike the skyrmion and 
in agreement with our earlier results, the field texture for the FV does not show any signature
of a domain wall separating unequivalent outer and inner regions. 

\section{\label{transitions}Field-driven transitions between skyrmionic and Vortical states}

In bulk and type II superconducting samples vortices with phase windings higher than 
$2n\pi$, where $n$ is integer, are energetically disfavored. 
The superconductor prefers two distant vortices each with phase winding $2\pi$ 
rather than one single vortex (giant vortex) with phase winding $4\pi$.
Nevertheless, in samples with dimensions of the order of the superconducting 
coherence length (mesoscopic samples), giant vortices can appear as 
stable configurations. The stabilization is provided mainly by the confinement
due to the small sample size, although the external magnetic field also contributes
through the screening currents and the confining force they exert on vortices. 
Field driven transitions from states with multiple distant vortices to giant 
vortices have been widely reported. \cite{gvs, Chibotaru_nat}

\begin{figure}[htb]
\includegraphics[trim={0.1cm 0.0cm 0.0cm 0.0cm}, clip=true, scale=1.45]{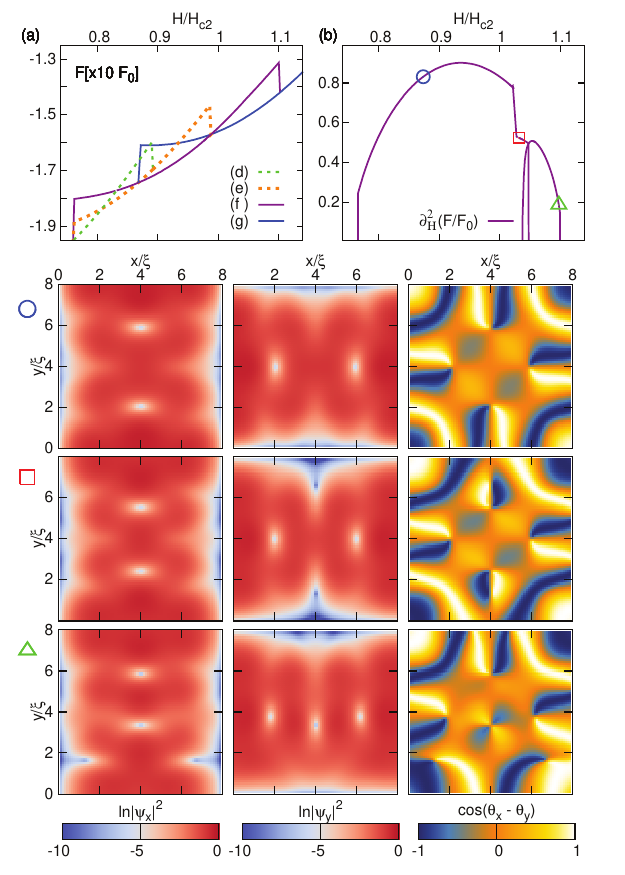}
\caption{(Color online) Field-driven transition from HQV to FV due to confinement in a square
mesoscopic sample of size 8$\xi\times$8$\xi$. (a) Energy  of the state \emph{f} 
of Fig. \ref{free_dk00} (a), along with some of its neighboring states.  
(b) Second order derivative of the energy with respect to the external field
showing three distinct states indicated by circular, squared and triangular 
symbols. The corresponding components of the superconducting order parameter are shown in 
panels (\textcolor{blue}{$\bigcirc$}), (\textcolor{red}{$\square$}) and 
(\textcolor{green}{$\triangle$}).
Displayed quantities are logarithmic contour plots of $|\psi_x|^2$ and $|\psi_y|^2$ 
in left and central columns, respectively, while the cosine of the phase difference
is shown at the right column.}
\label{trans_a}
\end{figure}

In this work we first report the field-driven transitions from HQV to FV states.  
Fig. \ref{trans_a} (a) shows the energy of state \emph{f} of Fig. \ref{free_dk00} (a),
along with some of its neighboring states. Panel (b) shows the second order derivative of the energy 
with respect to the external field only for state \emph{f} . While the energy of state \emph{f} is continuous, 
its second order derivative shows discontinuities indicating transitions between distinct states.
Three different states can be easily distinguished, which we labeled 
by a circle, square and triangle marker. The corresponding distributions of 
the superconducting order parameters are also shown in the figure: 
logarithmic contour plots of $|\psi_x|^2$ and
$|\psi_y|^2$ are shown in the left and central columns, while the cosine of 
the phase difference is shown in the right column. State (\textcolor{blue}{$\bigcirc$})
shows two fractional vortices in each component rendering four HQVs according to 
the phase difference contour plot. State (\textcolor{red}{$\square$}) shows two HQVs and two FVs. 
The FVs are composed of two fractional vortices belonging separately to each
component. The fractional vortices composing the FVs are slightly 
misaligned as can be seen in the density figures. This makes the FVs
display a small closed domain wall in the phase difference contour plot.
At high fields the screening currents confine even more the superconducting 
configuration of state (\textcolor{red}{$\square$}) transforming it into one state with 
three HQVs and one FV (\textcolor{green}{$\triangle$}). Due to the strong screening currents
the upper FV of state (\textcolor{red}{$\square$}) loses one of its fractional vortices 
which renders one HQV in state (\textcolor{green}{$\triangle$}). The strong confinement also forces 
the alignment of the fractional vortices composing the FV of state (\textcolor{green}{$\triangle$}).

\begin{figure}[htb]
\includegraphics[trim={0.1cm 0.0cm 0.0cm 0.0cm}, clip=true, scale=1.45]{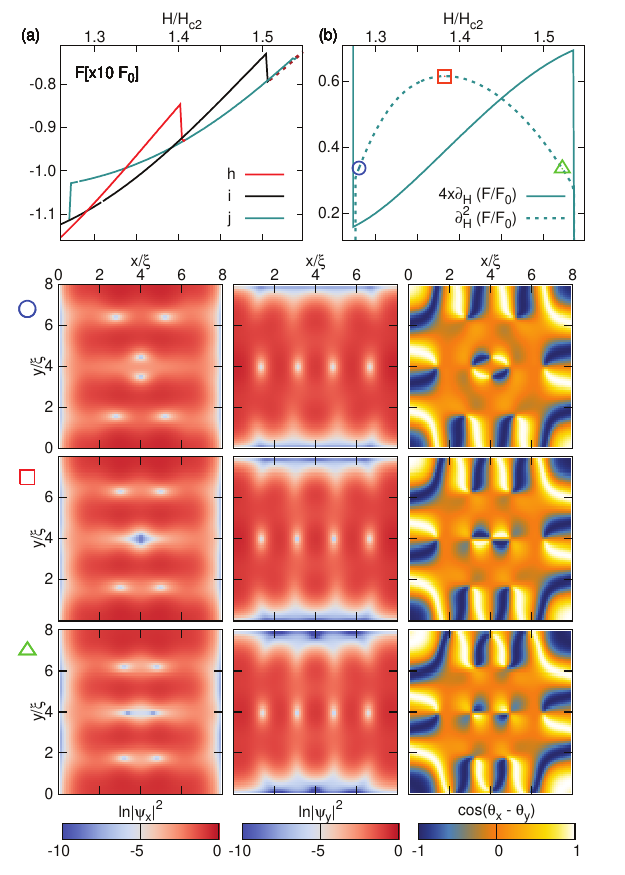}
 \caption{(Color online) Another example of a field-driven transition between skyrmionic and vortical states
 along the state \emph{j} of Fig.  \ref{free_dk00}.  Displayed quantities are the same as 
 in Fig. \ref{trans_a} with the only exception  that in panel (b)  the first order derivative
 of the energy with respect to the external field is also shown.}
\label{trans_b}
\end{figure}

Another field-driven transition from skyrmion to FV state  
is presented in Fig. \ref{trans_b}. Panel (a) shows the energy of state \emph{j}
of Fig. \ref{free_dk00}, along with some of its neighboring states.
Panel (b) shows  the first and second order derivatives of the energy 
with respect to the external field only for state \emph{j}. Unlike in Fig. \ref{trans_a} (b),
here the second derivative is continuous as well as the first derivative. 
Nevertheless, this does not mean that there are no distinct states along the 
stability curve of state \emph{j}. Circle, square and triangle markers 
(\textcolor{blue}{$\bigcirc$}, \textcolor{red}{$\square$} and \textcolor{green}{$\triangle$}) indicate
three states at weak, intermediate and strong confinement, respectively. 
At weak confinement the phase difference figure shows six HQVs and 
one skyrmion (see Fig. \ref{trans_b} (\textcolor{blue}{$\bigcirc$})). At intermediate confinement, 
state (\textcolor{red}{$\square$}) shows in $|\psi_x|^2$ that two out of the four fractional vortices 
composing the skyrmion of panel (\textcolor{blue}{$\bigcirc$}) have merged into one single discontinuity. 
This merger of initially distant fractional vortices renders the domain wall 
of the skyrmion asymmetric. At strong confinement (\textcolor{green}{$\triangle$}) the former fractional vortices
split their cores along the horizontal axis. According to the phase difference 
figure they join two other fractional vortices in density $|\psi_y|^2$
to form two horizontal FVs in the center of the sample. As can be easily seen, 
the vorticity of the superconducting components along this 
field-driven transition is constant, unlike in Fig. \ref{trans_a} where it was not.
This fact explains why the second order derivative is continuous here 
and discontinuous in Fig. \ref{trans_a}.

\section{\label{time_evol}Temporal Dynamic transitions}

To date, no works have treated the time-dependent phenomena within the GL formalism for 
chiral $p\,$-wave superconductors. Here we benefit from the temporal evolution included in the 
TDGL equations to report for the first time dynamic transitions involving vortices and skyrmions.

\begin{figure}[htb]
\includegraphics[trim={0.05cm 0.0cm 0.0cm 0.0cm}, clip=true, scale=1.45]{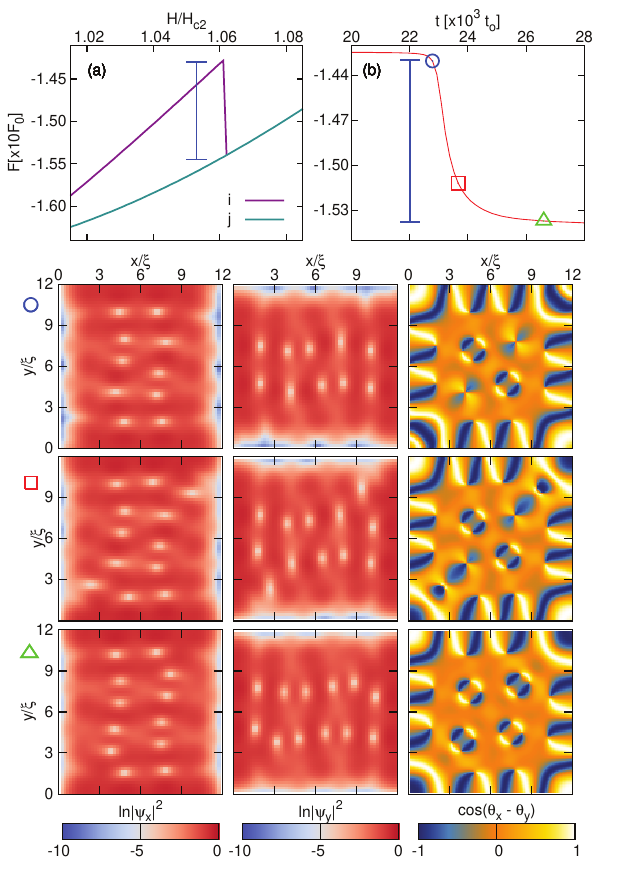}
 \caption{(Color online) Temporal vortex-skyrmion transition in a square mesoscopic sample of size 12$\xi\times$12$\xi$.
 Panel (a) shows the free energy of states \emph{i} and \emph{j} containing 10 and 12 fractional vortices per
 component, respectively. The energy of state \emph{i} is discontinuous at $H\approx 1.06H_{c2}$ reflecting a first order transition. 
 Panel (b) shows the temporal evolution of the energy at the latter transition. Three states, initial, intermediate and final
 are denoted by circle, square and triangle markers, respectively. The components of the superconducting
 order parameter  corresponding to each state are shown in  panels (\textcolor{blue}{$\bigcirc$}), 
 (\textcolor{red}{$\square$}) and (\textcolor{green}{$\triangle$}).}
\label{trans_temp}
\end{figure}

The dimensionless free energy as a function of the external field for states 
\emph{i} and \emph{j} is shown in panel (a) of Fig. \ref{trans_temp}. Unlike in Fig. \ref{free_dk00}
the sample size here is 12$\xi\times$12$\xi$ rather than the 8$\xi\times$8$\xi$, 
which was a suitable choice to study the evolution of the superconducting configuration. 
Panel (b) shows the temporal evolution of the free energy 
at the discontinuous step in energy in panel (a). Three states, initial, intermediate and 
final, are denoted by circle, square and triangle markers.
The corresponding superconducting order parameters are shown in 
panels (\textcolor{blue}{$\bigcirc$}), (\textcolor{red}{$\square$}) and (\textcolor{green}{$\triangle$}), respectively.
The displayed quantities in the latter panels are the same as in those of Fig. \ref{trans_a}. 
The initial state (\textcolor{blue}{$\bigcirc$}) is a multi-vortex-skyrmion state containing 
two pairs of skyrmions and FVs, surrounded by eight HQVs at the sample edges. 
This state was not obtained for sample size 8$\xi\times$8$\xi$ 
mainly due to the strong confinement there.
At the intermediate state (\textcolor{red}{$\square$}) two fractional vortices
nucleate in each component of the superconducting order parameter 
forming two FVs according to the phase difference contour plot. 
The four FVs of the intermediate state then
combine following the inverse process of the one described in Fig. \ref{trans_b},
to form two skyrmions as depicted in state (\textcolor{green}{$\triangle$}).
In the Supplementary Material, we present the animated data showing the temporal evolution 
of the superconducting order parameter, as a more convenient view of the transition
reported here. It is noteworthy here that all field-driven transition 
from HQV or skyrmion to FV states and vice-versa are essentially driven by HQV 
penetration and recombination into other topological entities.

\section{\label{aniso}Anisotropic case}

\subsection{Strong chiral limit}

This far, the ground states of a $p\,$-wave mesoscopic superconductor 
with size $8\xi\times 8\xi$ have been obtained under the assumption of weak coupling
and with a cylindrical Fermi surface, which led us to set the $k_i$ parameters to $1/3$.
However, several works have reported or suggested other scenarios for SRO
such as: (i) multiband superconductivity with the 1D Fermi sheets 
developing superconducting order, \cite{Deguchi, Garaudcoal} or
(ii) anisotropy in the cylindrical Fermi surface. \cite{Agterberg}
In order to include just anisotropy in the Fermi surface,
while preserving single-band superconductivity, 
in this section we introduce the parameter ($\delta_k$), which sets
the $k_i$'s  to: $k_1 = 1/3+2\delta_k$, and $k_2=k_3=1/3-\delta_k$.
The motivation behind this choice is that the theoretical values for the $k_i$ 
parameters corresponding to the three Fermi sheets ($\gamma$, $\alpha$ and $\beta$) lie between 
$1/3<k_i^\gamma\leq 1$ and $0 \leq k_i^{\alpha,\beta} <1/3$, \cite{Agterberg} respectively.
The GL equation for $p\,$-wave superconductors with anisotropy in the Fermi surface becomes:
 \begin{eqnarray}
 \partial_t\vec{\Psi}\!&=&\frac{2}{3}\!\Bigl[ \vec{D}^2\!+\!\Pi_+^2\hat{\sigma}_+\!+\! \Pi_-^2\hat{\sigma}_- \Bigr]\vec{\Psi} 
\!+\!\vec{\Psi}\Bigl(1\!-\!\frac{3|\vec{\Psi}|^2}{4}\!\pm\!\frac{\vec{\Psi}^*\hat{\sigma}_z\vec{\Psi}}{4}\Bigl) \nonumber \\ 
 &+& \delta_k\bigl[ \vec{D}^2 -2\bigl( \Pi_+^2\hat{\sigma}_+ + \Pi_-^2\hat{\sigma}_- \bigr)\bigr]\vec{\Psi}.
 \label{GL_par1}
 \end{eqnarray}
%
%meanwhile the superconducting current yields,+
%
%\begin{eqnarray}
% \vec{J} = {\rm Im}\biggl\{\Bigl(1/3 + \delta_k/2\Bigr)& &\vec{\Psi}^*\vec{D}\,\hat{\mathbb{I}}\,\vec{\Psi} \\
%+ \frac{1/3-\delta_k}{\sqrt{2}}& &\Bigl (\vec{\Psi}^*\Bigl[\Pi_+\hat{\sigma}_+ + \Pi_-\hat{\sigma}_-\Bigr]\vec{\Psi}\,\hat{\imath} \nonumber \\
%+&i&\,\vec{\Psi}^*\Bigl[\Pi_+\hat{\sigma}_+ - \Pi_-\hat{\sigma}_-\Bigr]\vec{\Psi}\,\hat{\jmath} \Bigr) \biggr\}. \nonumber 
%\label{sucurr_par1}
%\end{eqnarray}
 
By tuning $\delta_k$ within the interval $[0,1/3]$, the strength of the chiral terms is changed, 
therefore driving the system between two limiting cases: the left limiting case being at $\delta_k =0$ 
and given by Eq. (\ref{GL_cyl}), and the right limiting case being at $\delta_k = 1/3$ 
where the chiral coupling between the superconducting components is set to zero.

\begin{figure}[htb]
\center
\includegraphics[clip=true,trim={0.0cm 0.3cm 0.0cm 0.2cm}, scale=0.7]{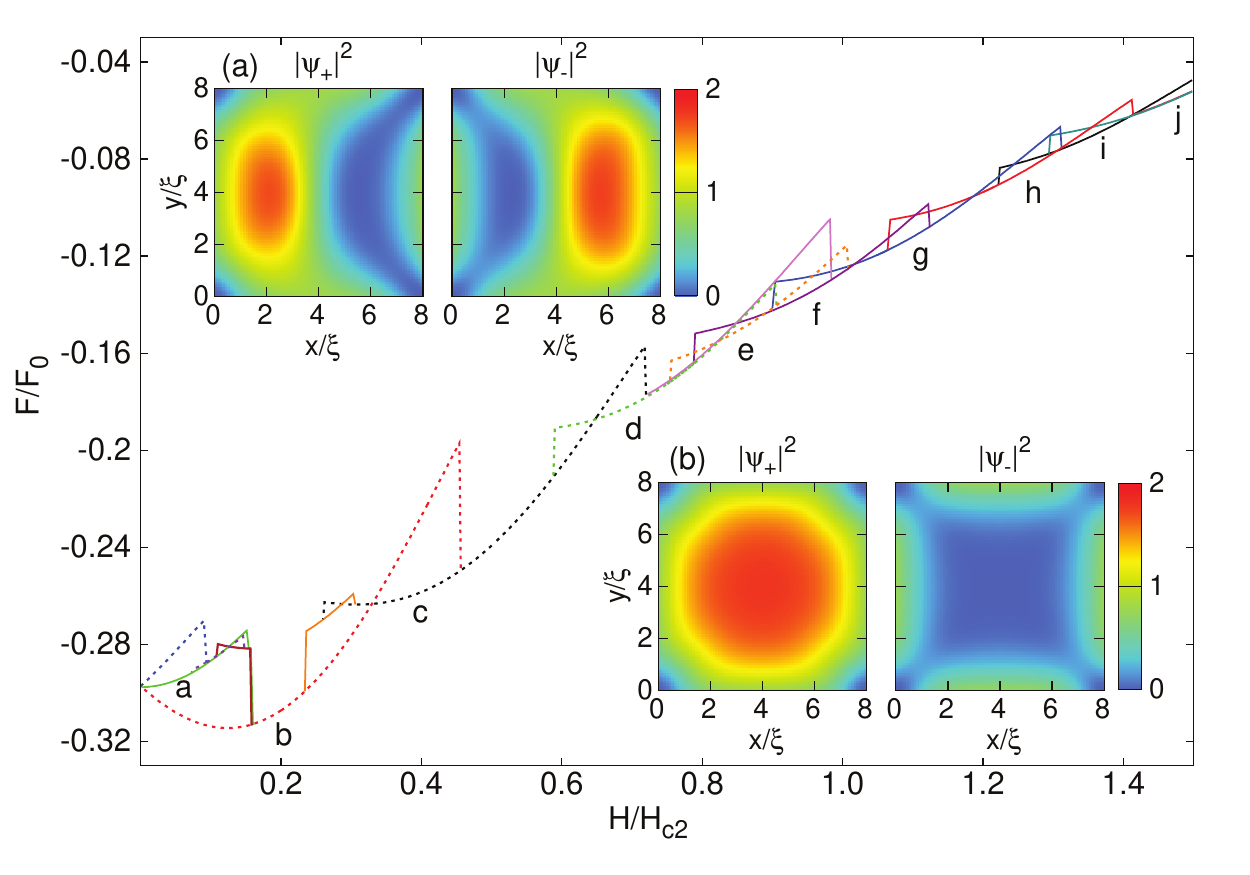}
\caption{(Color online) The free energy as a function of the external magnetic field, showing ground states \emph{b - j}  
plus one metastable state \emph{a}, from the numerical simulations using Eq. (\ref{GL_par1})
with $\delta _k = 0.03$. The parameters $k_i$ thus only slightly deviate from the value $1/3$
obtained when a cylindrical Fermi surface is considered. Panels (a) and (b) show the 
superconducting density components $|\psi_+|^2$ and $|\psi_-|^2$ of the states \emph{a} and 
\emph{b}, respectively.}
\label{free_dk30}
\end{figure}

Fig. \ref{free_dk30} summarizes the results obtained from the simulations that 
numerically approach Eq. (\ref{GL_par1}) with $\delta_k = 0.03$. The energy against field 
plot of Fig. \ref{free_dk30} shows nine ground states labeled by letters.   
The comparison between Figs. \ref{free_dk30} and \ref{free_dk00} reveals one 
important fact: the energy of the state \emph{a} is higher than the energy of its 
adjacent state \emph{b}. Actually, state \emph{a} here is no longer the ground state 
at low fields $H\approx 0$, unlike in Fig. \ref{free_dk00} where it was. Contour plots of the superconducting
order parameter ($\vec{\Psi}$) that correspond to the states \emph{a} and \emph{b} of Fig. \ref{free_dk30}
are depicted in insets (a) and (b), respectively. The comparison 
between the insets of Fig. \ref{free_dk30} and the corresponding states in Fig. \ref{st_dk00_pm} 
shows that despite of the small anisotropy introduced in the GL equation, 
the superconducting configuration of these states is practically identical in both cases. 

%\begin{figure}[htb]
%\includegraphics[trim={0.35cm 0.0cm 0.0cm 0.0cm}, clip=true, scale = 1.7]{tot_low_dk30}
%\includegraphics[trim={0.0cm 0.0cm 0.1cm 0.0cm}, clip=true, scale = 1.3]{psipm_low_dk30}
%\caption{Ground states (a-d) of Fig. \ref{free_dk30}. Displayed 
%quantities are the same as in Fig. \ref{st_dk00_pm}.}
%\label{st_dk30_pm}
%\end{figure}%

\begin{figure*}[htb]
\includegraphics[trim={0.2cm 0.4cm 0.0cm 0.1cm}, clip=true, scale = 1.45]{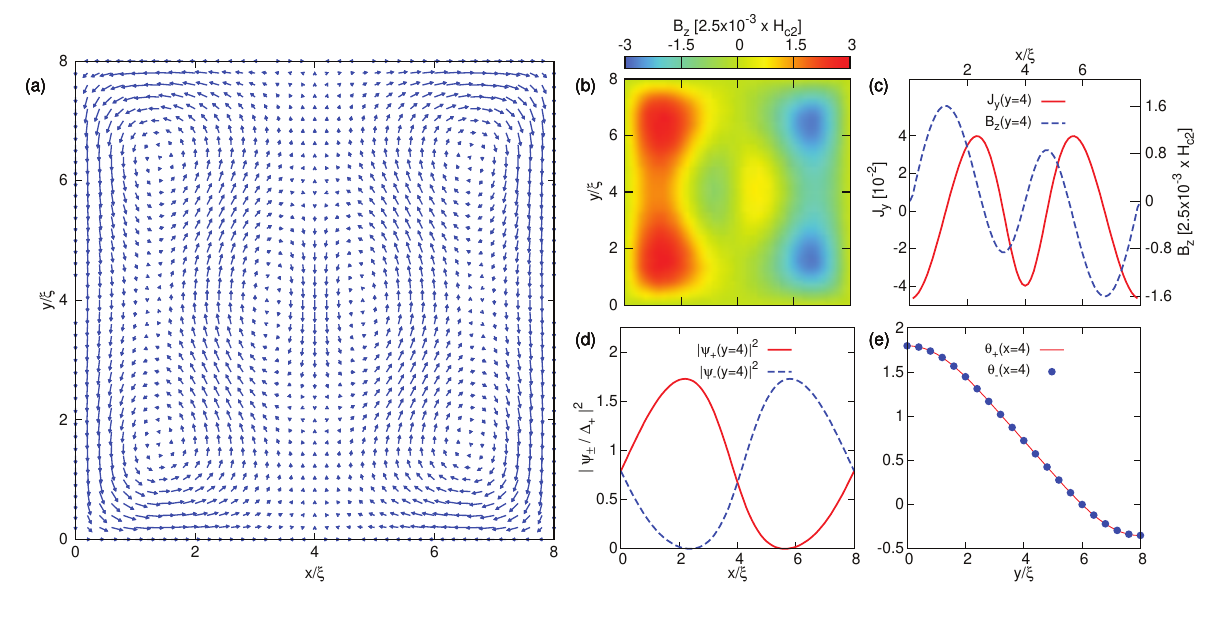}
\caption{(Color online) (a) Superconducting currents corresponding to the
state \emph{a} of Fig. \ref{free_dk30}. These currents, which 
were obtained at zero field, are composed of two edge currents with different chiralities and
flowing in opposite senses. (b) Contour plot of the magnetic induction ($B_z$) calculated
from the supercurrents of panel (a). (c) Line profiles of $J_y$ and $B_z$ along the line $y=4\xi$.
(d) Line profiles of $|\psi_\pm|^2$ corresponding to the state \emph{a} of Fig. \ref{free_dk30}.
(e) Line profiles of the angular phases of components $\psi_\pm$ along the line $x=4\xi$.}
\label{st_TRSB}
\end{figure*}

Two decades have passed since the discovery of the unconventional properties of strontium ruthenate, 
but to date there has not been a consensus whether or not it is a chiral $p\,$-wave
superconductor. \cite{Clifford, kirtleyone} The main experimental results that support 
unconventional superconductivity in SRO are provided by the set of measurements 
carried out using techniques such as the nuclear magnetic resonance (NMR), \cite{Ishida, Murakawa} 
$\mu$SR, \cite{Luke} the Kerr effect, \cite{Kapitulnik} and cantilever magnetometry. \cite{halfQVexp}  
The smoking gun evidence that lacks, and which, if found, would convince the scientific community is the finding 
of the theoretically predicted spontaneous currents in Sr$_2$RuO$_4$. \cite{Furusaki, Stone, Matsumoto}
Interestingly, what we just found in this work is that the state with spontaneous currents is no longer
the ground state when the GL model slightly deviates from the isotropic case at $H\approx 0$, 
i.e slightly deviated from the cylindrical Fermi surface. This energy lift of the 
state with spontaneous currents makes it even harder to be detected. Fig. \ref{st_TRSB}(a) 
shows the supercurrent distribution corresponding to the state \emph{a} of Fig. \ref{free_dk30}. 
We note that the currents displayed there were obtained at $H=0$, 
thus those are the spontaneous currents widely sought in experiments. 
The spontaneous currents are composed mainly of two counter-flowing streams at left and 
right sides of our sample. They are the chiral edge currents predicted by Furusaki \emph{et. al.}, \cite{Furusaki} Matsumoto
and Sigrist, \cite{Matsumoto} and Stone and Roy. \cite{Stone}  
Along the line $x=4\xi$, the linear domain wall (DW) of Fig. \ref{st_dk00_low}(a) separates the left and 
right sides showing an enhancement in the supercurrents around the center. The magnetic
induction corresponding to the supercurrents of panel (a) is shown in panel (b) of the same figure.

Panel (c) of Fig. \ref{st_TRSB} shows line profiles of the magnetic induction and the $y$-component
of $\vec{J}$ along the line $y=4\xi$. This plot agrees well with the result of Matsumoto and 
Sigrist which showed that $J_y$ ($B_z$) is an even (odd) function of $x$ along the line perpendicular
to the DW. \cite{Matsumoto} Finally, panels (d) and (e) provide important information that allow us
to calculate the supercurrent along the DW. From panel (d) the DW is defined by $|\psi_+|=|\psi_-|$
at $x=4\xi$, but along this line panel (e) tells us that not only the magnitudes of the 
superconducting components are equal but also their angular phases. Then, from Eq. (\ref{sucurr_gnral}) 
our estimation for the superconducting current along the linear domain wall is simply
$J_y(x=4\xi) = k_1|\psi_+|^2\partial_y\theta_+.$

\subsection{Strong Zeeman limit}

Microscopy with superconducting quantum interference devices (SQUIDs) and scanning Hall
probes (SHPs) have recently detected vortex coalescence in single crystals of strontium ruthenate.
\cite{Dolocan, Curran} One possible explanation for this behavior is the existence
of at least two different coherence lengths arising from multigap superconductivity, 
and which lead to attractive (repulsive) interaction at long (short) ranges.
\cite{Garaudcoal} Refs. [\onlinecite{kirtleyone}] and [\onlinecite{Curran}] have reported that within 
their corresponding resolutions no convincing evidence for spontaneous currents
and DWs has been found yet. In order to explore more superconducting configurations, 
comprising DWs, HQVs, FVs and skyrmions as the fundamental entities, 
in what follows a different set of the $k_i$ parameters is defined by: 
$k_1=1/3$, $k_2=1/3+\delta_k$ and $k_3=1/3-\delta_k$. Such a choice of parameters
enables one to keep constant the strength of the chiral terms while varying $\delta k$.
The first GL equation for this particular choice of parameters reads
 \begin{eqnarray}
 \partial_t\vec{\Psi} &=& \frac{2}{3} \Bigl[  \vec{D}^2 
  + \Pi_+^2\hat{\sigma}_+ + \Pi_-^2\hat{\sigma}_-  - \frac{3\delta_k}{2}  H \hat{\sigma}_z  \Bigr]\vec{\Psi} \nonumber \\ 
 &+& \vec{\Psi}\Bigl(1-\frac{3|\vec{\Psi}|^2}{4} \pm\frac{\vec{\Psi}^*\hat{\sigma}_z\vec{\Psi}}{4}\Bigl).
 \label{GL_par2}
 \end{eqnarray}

\begin{figure*}
\includegraphics[trim={0.0cm 0.1cm 0.0cm 0.2cm}, clip=true, scale=1.5]{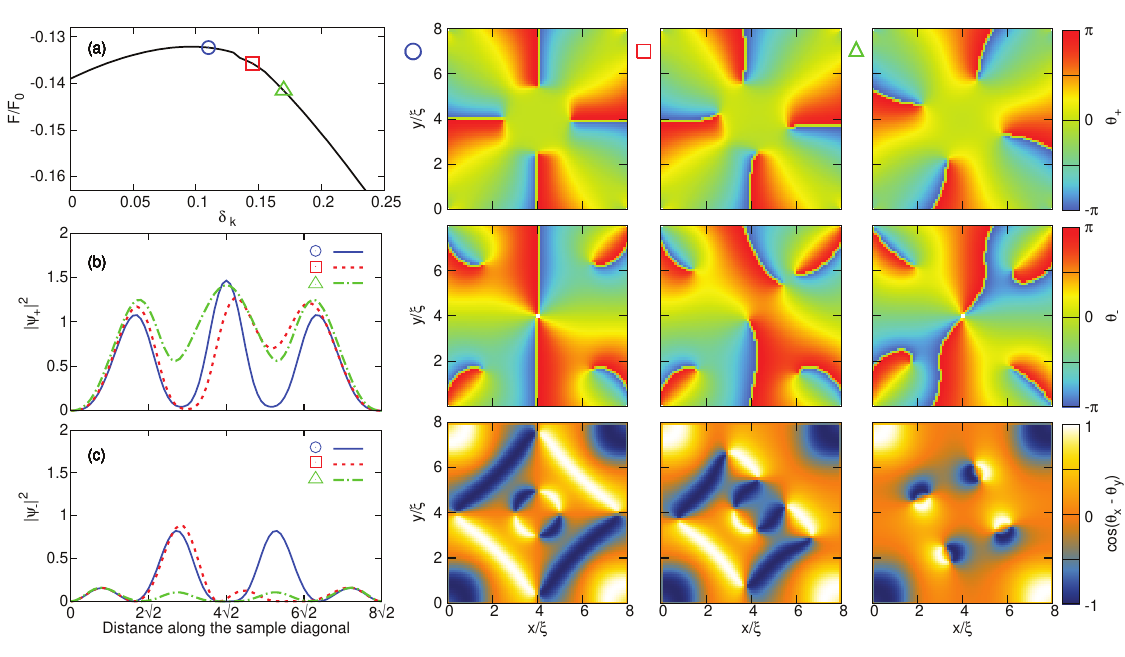}
\caption{(Color online) (a) Free energy as a function of the anisotropy parameter $\delta_k$
with the external magnetic field fixed at $H=0.530[H_{c2}]$. Three distinct 
states are indicated by circular, square and triangular markers
(\textcolor{blue}{$\bigcirc$}, \textcolor{red}{$\square$} and \textcolor{green}{$\triangle$}). 
(b) and (c) show line profiles of $|\psi_+|^2$ and $|\psi_-|^2$, respectively,
 along the diagonal line ($y=x$) corresponding to the states of panel (a).
Columns (\textcolor{blue}{$\bigcirc$}), (\textcolor{red}{$\square$}) and (\textcolor{green}{$\triangle$}) show contour plots
of $\theta_+$, $\theta_-$ and $\cos{(\theta_x-\theta_y)}$ corresponding to the denoted states of panel (a).
}
\label{st_vardk}
\end{figure*}

The resemblance of the fourth term in the upper line of Eq. (\ref{GL_par2}) 
with the Zeeman term leads us to interpret $\delta_k$ here 
as a sort of magnetic moment. In order to study the dependence of the 
superconducting configuration on the anisotropy parameter, in Eq. (\ref{GL_par2}) 
the magnetic field is kept fixed while $\delta_k$ is varied.  
Fig. \ref{st_vardk}(a) plots the free energy of the states, solving
Eq. (\ref{GL_par2}), as a function of the anisotropy parameter $\delta_k$. 
Circle, square and triangle markers denote three states whose 
$|\psi_+|^2$ and $|\psi_-|^2$ diagonal ($y=x$) line profiles are shown in panels (b) and (c), 
respectively. From panel (c), and unlike in panel (b),
one clearly sees that for high values of $\delta_k$ the density $|\psi_-|^2$
diminishes.  Our explanation for this behavior is through the definition 
of two effective coherence lengths, one for each superconducting component.
Defining them as the coefficients in front of the linear terms $\psi_+$ and
$\psi_-$ in Eq. (\ref{GL_par2}), they read: $\xi_+=1-\delta_k H$ and 
$\xi_-=1+\delta_k H$, respectively. With $H$ fixed and $\delta_k$ increasing,
$\xi_+$ ($\xi_-$) becomes smaller (larger) therefore leading to an effective 
reduction (increase) of confinement in component $\psi_+$ ($\psi_-$). 
Concerning the phase, contour plots of $\theta_+$, $\theta_-$ and 
$\cos{(\theta_x-\theta_y)}$ corresponding to the states denoted by circle, square and 
triangle markers are shown in rows $(\textcolor{blue}{\bigcirc})$, 
$(\textcolor{red}{\square})$, and $(\textcolor{green}{\triangle})$. According to the phase difference figure, 
state (\textcolor{blue}{$\bigcirc$}) is composed of two concentric skyrmions, 
one circular and one rhomboidal. From the contour plot of $\theta_-$ 
one sees that the circular skyrmion arises from the formation of 
one giant vortex in $\psi_-$ with phase winding $4\pi$. The phase difference figure
corresponding to state (\textcolor{red}{$\square$}) shows one irregular closed domain wall
emerging from the intersection of the circular and rhomboidal skyrmions.
Its formation is determined by the annihilation of the giant vortex in 
$\theta_-$ that has split into two fractional vortices. Finally,
the phase difference figure of state (\textcolor{green}{$\triangle$}) shows four FVs 
with cores slightly asymmetric as can be seen from the small circular
DWs present there. Due to the density $|\psi_-|^2$ has been substantially
depleted at this value of $\delta_k$, the superconducting state is 
completely defined by component $\psi_+$. Hence, what we have achieved
by adding anisotropy to the chiral $p\,$-wave model of Eq. (\ref{GL_gnral}),
is a chiral polarization enhanced due to the strong confinement
present in a mesoscopic sample.

\section{\label{summary}Conclusions}

In summary, we have studied in detail the Ginzburg-Landau model that describes 
$p\,$-wave tetragonal superconductors \cite{Sigrist,Agterberg} and  all the possible states of 
a mesoscopic superconducting sample as a function of the external magnetic field and the anisotropy
parameters of the material. Due to the odd parity the order parameter is a two-component 
complex vector and the fundamental solutions of the corresponding TDGL equations,
that we obtained numerically, are fractional vortices, i.e. solutions where 
the phase winding $2\pi$ is found in one component but not in the other one. 
In two- and three-band superconductors similar fractional vortices were obtained
between components, but for different reasons. \cite{Chibotaru2010,Geurts2010,gillis}  
Fractional vortices in different components can combine to form a cored/full-vortex state, 
as well as a coreless/skyrmion state seen in phase difference and magnetic response figures. 
Skyrmions arise when same number of fractional vortices in each component combine to form  
a closed domain wall that separates distinct intercomponent phase difference ($\theta_x-\theta_y$)
regions. \cite{Garaudprl,Garaudech} Alternating segments between $1$ and $-1$ in  the $\cos{(\theta_x-\theta_y)}$ 
between fractional vortices along the domain wall is the main signature for skyrmions. 
While for skyrmions the topological charge 
($\mathbb{Q}$) is defined by the hopf invariant, \cite{expskyone,Garaud} 
for vortices it is by the circulation of the superconducting velocity. 
The screening currents of half-quantum vortices are anisotropic and in Cartesian 
coordinates we have analytically shown that the equipotential lines of the screening currents
are ellipsoidal rather than circular as in full vortices. 
This anisotropic screening leads half quantum vortex to be attracted towards the sample edges. 
The mesoscopic size of our samples provides stability 
to the half quantum vortices and the $\mathbb{Q}\!=\!2$ skyrmions, 
in contrast to larger systems where larger values of $\mathbb{Q}$ were considered, \cite{Garaud}  
and bulk systems where the half quantum vortices have been usually regarded as 
high-energy states. \cite{halfQV,Babaevprl} Actually the mesoscopic size of the sample plays 
a remarkable role in the stability of skyrmions as well as in the here reported novel transitions
(e.g. from a skyrmion to a full vortex). At high external fields, above the $H_{c1}$ critical
field, states with different configurations of skyrmions and half quantum vortices gradually transform into 
full vortices owing to the increased screening currents and confinement effects.

To date, the only superconductor expected to be $p\,$-wave is strontium ruthenate, with enough evidence 
demonstrating its unconventional behavior. \cite{Luke,Ishida,Murakawa,Kapitulnik,halfQVexp} 
Nevertheless, many works have failed to convincingly detect spontaneous currents, 
half-quantum vortices and skyrmions in large samples. What we have demonstrated here is that: 
(i) even by slight anisotropy in the Fermi surface, the state with spontaneous currents 
is no longer the ground state at $H\approx 0$, and (ii) field-driven transitions between 
half-quantum vortex, full vortex, and skyrmions provide 
a better method to indirectly prove their existence in magnetic measurements.

% Surround figure environment with turnpage environment for landscape
% figure
% \begin{figure}
% \end{figure}

% tables should appear as floats within the text
% The ruledtabular enviroment adds doubled rules to table and sets a
% reasonable default table settings.
% Use the table* environment to get a full-width table in two-column
% Add \usepackage{longtable} and the longtable (or longtable*}
% environment for nicely formatted long tables. Or use the the [H]
% placement option to break a long table (with less control than
% \begin{table}%[H] add [H] placement to break table across pages
% \begin{ruledtabular}
% Lines of table here ending with \\
% \end{tabular}
% \end{ruledtabular}
% \end{table}

% Surround table environment with turnpage environment for landscape
% table
% \begin{table}
% \begin{ruledtabular}
% \end{tabular}
% \end{ruledtabular}
% \end{table}

%\appendix
%\section{}

\begin{acknowledgments}
This work was supported by the Research Foundation - Flanders (FWO). 
E.S. acknowledges support from the Sao Paulo Research Foundation (FAPESP). 
\end{acknowledgments}


\end{document}